\documentclass[aps,twocolumn]{revtex4}
\usepackage{graphics}
\begin{document}
\newcommand{\bstfile}{aps} 
\draft
\title{Two-photon fluorescence measurements of reversible photodegradation in a dye-doped polymer}
\author{Ye Zhu, Juefei Zhou, and Mark G. Kuzyk}
\address{Department of Physics and Astronomy, Washington State University, Pullman, Washington 99164-2814}
\date{\today}

\begin{abstract}
We report on the dynamics of photodegradation and subsequent recovery of two-photon fluorescence in a dye-doped polymer.  The energy dependence suggests that photo-degradation is a linear process while recovery is entropic.   Such recovery could be useful to high-intensity devices such as two-photon absorbers, which can be used in many applications. 
\end{abstract}

\maketitle

OCIS:  160.2540, 160.4330, 160.4890, 160.5470, 190.4180, 140.3330

\vspace{1em}

\begin{figure}
\includegraphics{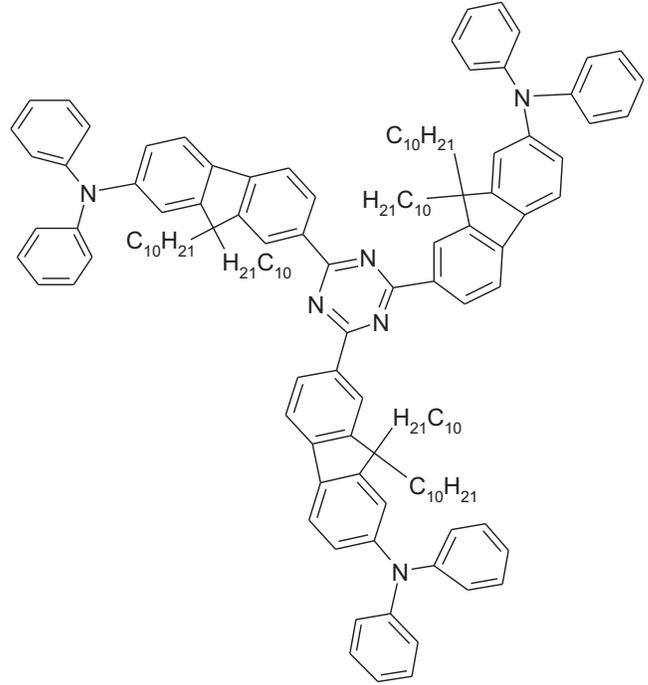}
\caption{The AF455 molecule.} \label{fig:molecule}
\end{figure}

The problem of photodegradation of dye molecules is well known to high-intensity applications such as in gain media and laser media.\cite{dyuma92.01,popov.98.01}  Peng and coworkers showed that the fluorescence spectrum from rhodamine B-doped in a PMMA fiber decreased as a function of time when exposed to an intense laser source but that the florescence signal would partially recover.\cite{Peng98.01}  Howell and coworkers found that the amplified spontaneous emission (ASE) signal of DO11 dye-doped PMMA polymer would fully recover when left in the dark for a couple of days;\cite{howel02.01} but in liquids, there was no recovery.\cite{howel04.01} (In the liquid studies, the total sample volume was exposed to prevent recovery due to mass transport from the surrounding reservoir.) In the solid solution experiments, the decay constant increased and the recovered ASE efficiency increased with subsequent cycling of degradation and recovery - suggesting that it may be possible to harden a material against photodegradation by such cycling.  The recovery mechanisms was attributed to phototautomerization followed by dimer formation.\cite{kuzyk06.06} In the present work, we report on photodegradation and recovery of two-photon fluorescence (TPF) in the chromophore AF455 (shown in Figure \ref{fig:molecule}) doped in poly (methyl methacrylate) (PMMA) polymer.  AF455 is known to have a large two-photon absorption cross-section and is therefore a promising optical material for many applications.\cite{kanna04.01}

\begin{figure}
\includegraphics{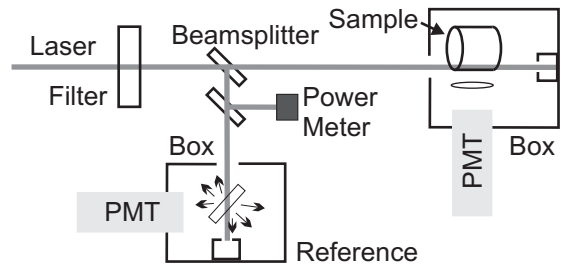}
\caption{Experimental layout.  The beam intensity is adjusted with a half-wave-plate/polarizer pair (not shown).} \label{fig:experiment}
\end{figure}

Figure \ref{fig:experiment} shows the experimental layout.  The laser beam is filtered to remove all parasitic light and is split into two parts.  Each beam passes into a light-tight black box through a small aperture.  The beam scatters from a piece of glass in the reference arm and is monitored with a photomultiplier tube (PMT).  The surface of an AF-455 dye-doped PMMA polymer cylinder (with doping approximately 0.06\% by weight) is illuminated in the sample arm and the TPF signal is collected by a lens and imaged onto the photocathode of a PMT.  The response of the PMT is spectrally broad enough to encompass the whole TPF spectrum, so, the measured signal is proportional to the integrated TPF energy.  A small portion of the incident beam is deflected to a power meter to allow for continuous monitoring of long-term drift in laser power while the reference arm is used to account for pulse-to-pulse fluctuations.  The PMT signals are read by a digital oscilloscope and stored for subsequent analysis. 

\begin{figure}
\includegraphics{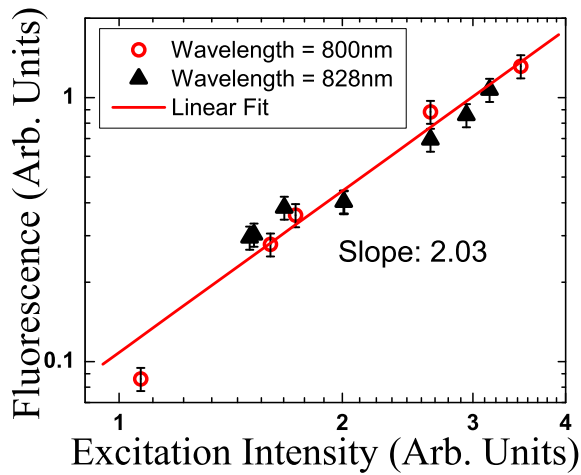}
\caption{Two-photon fluorescence as a function of pump intensity.} \label{fig:PowerLaw}
\end{figure}

The laser source is a Continuum tunable OPG laser pumped with tripled Nd:YAG.  The pulses are about 25ps in duration at 10Hz.  Since the AF455 sample is known to have higher two-photon fluorescence in the range 790-830nm, the experiment was carried out at two wavelengths: 800nm, where TPF during degradation and recovery were measured for laser energies of $6$, $7$, and $9 \, \mu J$/pulse; and, at 828nm, where laser energies of $3.5$ and $7 \, \mu J$/pulse were used.    The $7 \, \mu J$/pulse runs at the two wavelengths were identical within experimental uncertainty.  Note that the peak power of the $9 \, \mu J$ pulse is about $0.36 \, MW$.  About half of the laser energy reaches the sample, yielding energies of $1.75$, $3$, $3.5$, and $4.5 \, \mu J/pulse$.  Figure \ref{fig:PowerLaw} shows that the signal is quadratic in the pump intensity at the two measurement wavelengths, as required by two-photon absorption.  Comparing our signal with the signal from a DCM laser dye as a standard - as determined by Xu and Webb,\cite{xu96.02} the two photon cross-sections at 800nm and 828nm are determined by an independent experiment to be $90 (\pm 20) \, 10^{-50} cm^4 s/photon $ and $167 (\pm 35) \, 10^{-50} cm^4 s/photon $, respectively.

The sample was exposed to the laser source for 3 hours, and the signals were recorded with the oscilloscope every 5 minutes.  The laser beam was then blocked to allow the sample to recover.  A reading was taken every half hour for 30 seconds over an 8 hour recovery period.  The sample was kept in the dark between readings so that over the eight-hour period, the sample was exposed to the pump laser for only 4 minutes over which time at most $11 \, mJ$ of energy was deposited.  After the recovery experiment is completed, the sample is left in the dark box for 3 days so that it can fully recover before the next experiment. 

\begin{figure}
\includegraphics{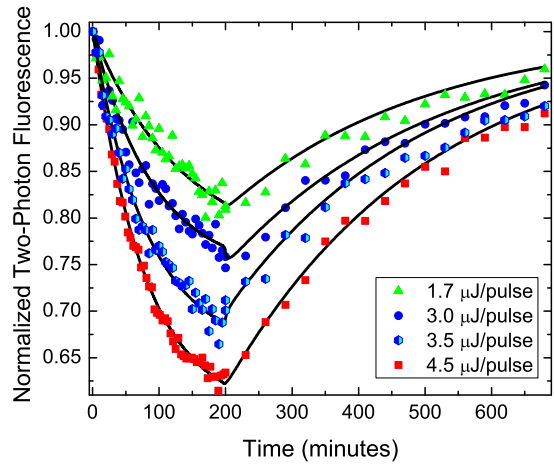}
\caption{Decay and recovery of TPF when pumped at four different pump intensities.} \label{fig:Dynamics}
\end{figure}

The points in Figure \ref{fig:Dynamics} shows the TPF signal, normalized to the reference signal, as a function of time, for four different pump pulse energies.  While the system is being pumped, the TPF signal decays, implying a degradation process that depends on intensity.  When the pump is turned off, the signal recovers, implying an entropic process that opposes the degradation process.

To model this behavior, we assume that in a time interval $dt$ a molecule absorbs $\alpha' I dt$ photons where $I$ is the number of photons per second incident on the molecule and $\alpha'$ the fraction of photons absorbed.  If the fraction of each absorbed photon that results in degradation is $\gamma$, and the entropic decay rate is $\beta$, then the change in the number of fluorescing molecules, $dN$  in the time interval $dt$ is,
\begin{equation}\label{diffeqq}
dN = -N \alpha I dt + \beta \left( N_0 - N \right) dt,
\end{equation}
where $N_0$ is the initial population of two-photon fluorescing molecules, $\alpha \equiv \gamma \alpha'$, and $N_0 - N$ is the population of non-fluorescing molecules. 

The solution to Equation \ref{diffeqq} for the degradation process is 
\begin{equation}\label{nDecay}
n = \frac {\beta} {\beta + \alpha I} + \frac {\alpha I } {\beta + \alpha I} \cdot e^{- \left( \beta + \alpha I \right) t },
\end{equation}
where we define $n$ = $N/N_0$, and the recovery process is given by
\begin{equation}\label{nRecover}
n = 1 - \left( 1-n(t_0) \right) e^{- \beta t },
\end{equation}
where $t_0$ is the time at which the pump laser is turned off and $n(t_0)$ is the population of molecules given by Equation \ref{nDecay} at that time.  Since the TPF signal should be proportional to the number of chromophores, $N$, the TPF signal normalized to unity at time $t=0$ should also follow Equation \ref{nDecay}. The smooth curves in Figure \ref{fig:Dynamics} represent a fit of the data to Equations \ref{nDecay} and \ref{nRecover} with $\alpha I$ and $\beta$ as the two adjustable parameters..

\begin{figure}
\includegraphics{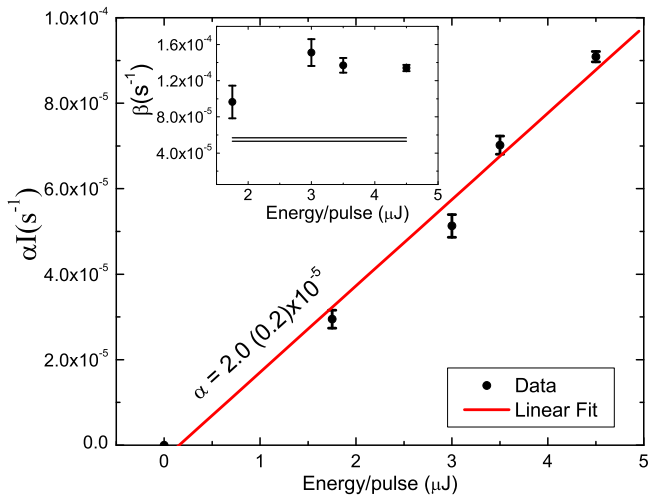}
\caption{$\alpha I$ and $\beta$ as a function of laser pulse energy as determined from the fit to Equation \ref{nDecay} as shown in Figure \ref{fig:Dynamics}.  The horizontal lines in the inset represent the range of $\beta$ values determined from the recovery process according to Equation \ref{nRecover}.} \label{fig:alpha}
\end{figure}
We first consider the time dependence of TPF recovery.  According to Equation \ref{nRecover}, the time constant should be independent of the laser power used in the degradation process.  Indeed, we find the time constants to all be the same within experimental uncertainty.  The average is $\beta = 5.51 (\pm 0.16) \times 10^{-5} \, s^{-1}$.  Figure \ref{fig:alpha} shows the values of $\alpha I$ and $\beta$ that are determined from a two-parameter fit of the decay data given by Equation \ref{nDecay}.  First, we note that $\alpha I$ is a linear function of the laser pulse energy.  The degradation mechanism therefore must be a linear function of the intensity for the range of measured intensities, suggesting the mechanism to be a linear process such as one-photon absorption.  The inset shows the range of $\beta$ values determined from the recovery run using Equation \ref{nRecover} (two horizonal lines) as well as the from the decay run using Equation \ref{nDecay} (points).  In the limit of low intensity, the two appear to converge.  So, we propose the hypothesis that during degradation, heating of the sample can accelerate the recovery process.  This suggests that the recovery mechanism is entropic in nature.

In conclusion, we have observed that photodegradation of a solid solution of the octupolar AF-455 chromophore doped in PMMA polymer is reversible when the sample is placed in the dark for over 9 hours.  The TPF signal as a function of time is well characterized by a model that takes into account the competition between photodegradation and entropic recovery.  The energy-dependence of the degradation process suggests that a linear process such as one-photon absorption is responsible.  The time dependence of TPF during recovery for each run are single exponentials, all with the same time constant, which is consistent with our assumption that the recovery process is entropic in nature.  Furthermore, the low-intensity limit of the degradation data appears to give the same recovery time constant while at higher energies, where the sample absorbs more energy and is therefore heated, the time constant is shorter.  This is also consistent with the entropic model of recovery.

In high-intensity applications where photodegradation is a hurdle, understanding the mechanisms of the recovery process may help in the design of better materials.  Future studies will be aimed at higher intensities, where degradation is dominated by multi-photon absorption.  Simultaneous measurements of linear and nonlinear spectroscopy along with TPF will undoubtedly lead to a better understanding of the underlying recovery process, which in turn, would enable the design of hardened materials.

{\bf Acknowledgements: } We thank Dr. Shaoping Bian for making the samples and acknowledge the National Science Foundation (ECS-0354736) for generously supporting this work.  The Air Force Research Laboratory, Materials and Manufacturing Directorate supplied materials and generously supporting this work.  


\clearpage

\end{document}